\documentclass[aps,pra,reprint,superscriptaddress]{revtex4-2}

\usepackage{amsmath,amsfonts,amssymb}
\usepackage{xcolor,graphicx}
\usepackage{tikz}
\usepackage[pdftex]{hyperref} 
\usepackage{braket}
\usepackage{dcolumn}
\usepackage{bm}
\hypersetup{
	colorlinks = true,
	linkcolor = blue,
	anchorcolor = blue,
	citecolor = red,
	filecolor = blue,
	urlcolor = blue}

\usepackage{float}
\begin{document}

\title{Effects of squeezing on the power broadening and shifts of micromaser lineshapes}

\author{L. Hernández-Sánchez}%
\email[e-mail: ]{leonardi1469@gmail.com}
\affiliation{Instituto Nacional de Astrofísica Óptica y Electrónica, Calle Luis Enrique Erro No. 1\\ Santa María Tonantzintla, Puebla, 72840, Mexico}
\author{I. Ramos-Prieto}%
\affiliation{Instituto Nacional de Astrofísica Óptica y Electrónica, Calle Luis Enrique Erro No. 1\\ Santa María Tonantzintla, Puebla, 72840, Mexico}
\author{F. Soto-Eguibar}%
\affiliation{Instituto Nacional de Astrofísica Óptica y Electrónica, Calle Luis Enrique Erro No. 1\\ Santa María Tonantzintla, Puebla, 72840, Mexico}
\author{H.M. Moya-Cessa}%
\affiliation{Instituto Nacional de Astrofísica Óptica y Electrónica, Calle Luis Enrique Erro No. 1\\ Santa María Tonantzintla, Puebla, 72840, Mexico}
\date{\today}
\begin{abstract}
It is well known that AC Stark shifts have an impact on the dynamics of atoms interacting with a near-resonant quantized single-mode cavity field, which is relevant for single-atom micromasers. In this study, we demonstrate that when the field is in a squeezed coherent state, the micromaser lines are highly sensitive to the squeezing parameter. Furthermore, we show that when considering a superposition of squeezed coherent states with equal amplitude, the displacement of the transition lines depends significantly not only on the squeezing parameter but also on its sign.
\end{abstract}
\maketitle
\section{Introduction}
Major steps towards the generation of nonclassical states of the quantized electromagnetic field have been taken over the years.  Nonclassical states usually exhibit less fluctuations or noise than coherent states~\cite{Glauber_1963a,Glauber_1963b} for certain observables; for this reason, noise associated to coherent states is referred to as the standard quantum limit. Among the nonclassical states that have attracted much interest over the years are squeezed states~\cite{Yuen_1979,Caves_1981,Loudon_1987}, states that are known to produce notable effects in the atomic inversion of a two level atom~\cite{Satyanarayana_1989,Moya_1992}, and in the resonance fluorescence spectrum~\cite{Villanueva_2020a} when they interact with a single two-level atom~\cite{Fu_2021}. This is because the shape of the photon distribution (that may be very different from that of a coherent state)  it is directly reflected in both, the atomic inversion and the atomic spectrum; such photon distribution imprints it signature in the atomic inversion, producing the so-called \textquotedblleft ringing revivals\textquotedblright \;~\cite{Satyanarayana_1989,Moya_1992}. Squeezed states are also  of great importance in cases where the detection of light needs to be extremely efficient, such as gravitational wave detectors; in fact, gravitational wave interferometers obtain their great sensitivity by combining suspended masses and squeezed  states of light~\cite{Tse_2019,McCuller_2021,Chua_2014}.\\

On the other hand, in experimental studies of micromasers~\cite{Filipowicz_1986}, asymmetries and Stark shifts in lineshapes have been observed and attributed to the effects of nearby off-resonant levels~\cite{Meschede_1985}. Such lineshapes are affected by the photon distribution of the quantized field; in particular, for a number state, the AC Stark shift moves the power-broadened resonance away from its unperturbed value; however, for a field with many contributions in the photon number, the distinct distribution of shifts contributes to a mean shift and to an effective broadening of the resonance~\cite{Moya_1991}. The micromaser lineshapes are very sensitive to the temperatures at which the micromaser is operated~\cite{Rempe_1990a,Rempe_1990b} and at lower temperatures thermal asymmetries become much less significant~\cite{Rempe_1989}.\\

The effects of nearby levels can also be observed in the resonance fluorescence spectrum, where the Stark shift created by such levels produces displacements of the resonance fluorescence peaks, because of an induced effective detuning~\cite{Villanueva_2020b,Zhang_2007}. In fact, it has been shown that the AC Stark effect can influence the transfer of quantum entangled information~\cite{Wang_2008}. Besides, the Doppler-free absorptive lineshape has been formulated  for coherently prepared drive-probe systems, where the individual contributions of drive and probe fields on AC Stark splitting could be demonstrated by the broadening of lineshapes~\cite{Ghosh_2015}. Comparison between the AC Stark effects and effects produced by nonlinear interactions, namely, the Kerr interaction, has been carried out~\cite{Moya_1995a,Berlin_2001}.\\

In this work, we address the study of the Jaynes-Cummings model with the AC Stark term, taking into account nearby non-resonant levels. In Section~\ref{Sec1}, we solve the Schr\"odinger equation and examine specific cases in which the atom is initially in either its first excited state or its ground state. Using a squeezed coherent state as the initial condition for the field, in Section~\ref{Comprimidos}, we analyze the effects of nearby non-resonant levels on atomic inversion and demonstrate how the lineshapes are distorted and broadened as the squeezing parameter $r$ varies. In Section~\ref{Superposición}, we extend the analysis to a superposition of squeezed coherent states with the same atomic conditions as in Section~\ref{Comprimidos}. In this section, we show that the changes in the lineshapes are highly sensitive, not only to the amplitude of the squeezing parameter, but also to its sign. Finally, in Section~\ref{Conclusión}, we present our conclusions.

\section{ac Stark shift Hamiltonian}\label{Sec1}
\begin{figure}[H]
\centering
\includegraphics{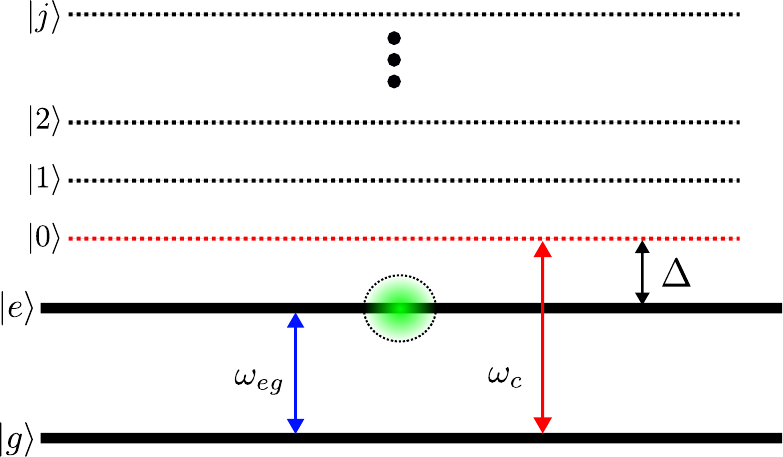}
\caption{Energy level diagram showing the pair of nearly resonant excited atomic states with transition frequency $\omega_{eg}$, the field frequency $\omega_{c}$, and a set of non-resonant levels that participate only virtually in the excitation and are responsible for the AC Stark shifts at the transition frequency $\omega_{eg}$.}
\label{fig1}
\end{figure}

Let us consider an atom with a ground state $\ket{g}$, an excited state $\ket{e}$, and higher states denoted by $\ket{j}$, with $j = 0, 1, 2, \ldots$ The atom interacts with a single-mode field, as shown in Fig.~\ref{fig1}. We consider that the field is approximately tuned to the transition frequency between the levels $\ket{g}$ and $\ket{e}$ of the atom but detuned from the nearby levels $\ket{j}$ (AC Stark effect). The Hamiltonian that describes this system is expressed as~\cite{Meschede_1985,Moya_1991,Moya_1995b,Villanueva_2020b,Hernandez_2023}
\begin{equation} \label{eq:1}
\hat{H} = \frac{\omega_{eg}}{2}\hat{\sigma}_{z} + \omega_c\hat{a}^\dagger\hat{a}+ \chi\hat{a}^\dagger\hat{a}\hat{\sigma}_{z}+ g\left(\hat{\sigma}_{+}\hat{a} + \hat{\sigma}_{-}\hat{a}^\dagger\right),
\end{equation}
where $g$ is the coupling constant between the two-level system and the field (in the dipole approximation), while $\chi$ is the parameter that quantifies the intensity of the interaction in the AC Stark effect, due to the presence of nearby non-resonant virtual levels. Creation and annihilation operators, $\hat{a}^\dagger$ and $\hat{a}$, are used, which satisfy the commutation relation $\left[\hat{a}, \hat{a}^\dagger\right] = 1$. Additionally, to describe the atomic part of the system, we use the operators $\hat{\sigma}_{+} = \ket{e}\bra{g}$, $\hat{\sigma}_{-} = \ket{g}\bra{e}$, and $\hat{\sigma}_{z} = \ket{e}\bra{e} - \ket{g}\bra{g}$, which satisfy the commutation relations $[\hat{\sigma}_{+}, \hat{\sigma}_{-}] = \hat{\sigma}_{z}$ and $[\hat{\sigma}_{z}, \hat{\sigma}_{\pm}] = \pm 2 \hat{\sigma}_{\pm}$.
We proceed to an interaction picture, moving to a frame that rotates at frequency $\omega_c$; i.e., by means of the time dependent unitary transformation $\hat{\mathcal{R}}=\exp\left[\mathrm{i} \omega_c t (\hat{n}+ \hat{\sigma}_z /2)\right]$, to produce the Schrödinger equation
\begin{equation}\label{Schrodinger}
    \mathrm{i} \frac{d \ket{\Psi (t)}}{dt} = \hat{\mathcal{H}} \ket{\Psi (t)},
\end{equation}
with the interaction Hamiltonian
\begin{equation}\label{Hamiltoniano2}
\begin{split}
\hat{\mathcal{H}} &= \hat{\mathcal{R}}\hat{H} \hat{\mathcal{R}}^{\dagger}-\mathrm{i}\hat{\mathcal{R}}\partial_t\hat{\mathcal{R}}^\dagger,\\
&=\left( \frac{\Delta}{2} + \chi \hat{n} \right) \hat{\sigma}_z + g \left( \hat{\sigma}_{+}\hat{a} + \hat{\sigma}_{-} \hat{a}^{\dagger} \right),
\end{split}
\end{equation}
being $\Delta = \omega_{eg} - \omega_c$ the detuning between the field frequency and the atomic transition frequency.\\
From the Hamiltonian $\hat{\mathcal{H}}$, the dynamics of any initial condition can be obtained by solving the Schrödinger equation~\eqref{Schrodinger} by one of the known methods~\cite{Klimov_2009,Amaro_2015}. In this work, we use the traditional method that proposes at time $t$, the atom-field state vector as a superposition of Fock states $\left\{\ket{n}\right\}$. Given that only two (composed) levels are coupled by the above Hamiltonian, namely $ \ket{n} \ket{e}$ and $\ket{n+1} \ket{g}$, we may write the solution as
\begin{equation}\label{Solucion}
\ket{ \Psi (t) }=\sum_{n=0}^{\infty}\left[C_n (t) \ket{n} \ket{e}+D_n(t) \ket{n+1} \ket{g} \right],
\end{equation}
where the coefficients $C_n(t)$ and $D_n(t)$ are to be determined; in order to do so, we insert this proposal into the Schrödinger equation~(\eqref{Schrodinger}), and the problem is reduced to solve the system of coupled differential equations
\begin{align}
\mathrm{i}\frac{d}{dt}
    \begin{bmatrix}
        C_n(t)\\D_n(t)\\
    \end{bmatrix}
    =& 
    \begin{bmatrix}
        \chi n +\frac{\Delta}{2}&g\sqrt{n+1}\\
        g\sqrt{n+1}&-\chi(n+1)-\frac{\Delta}{2}\\
    \end{bmatrix}
        \begin{bmatrix}
        C_n(t)\\D_n(t)\\
    \end{bmatrix},
\nonumber \\ &
 n=0,1,2,\dots.
\end{align}
The general solution of these differential equations is
\begin{align}\label{0050}
    \begin{bmatrix}
        C_{n}(t)\\D_n(t)
    \end{bmatrix}
    =&\exp\left(\mathrm{i}  \frac{\chi t}{2}\right)
    \begin{bmatrix}
        M_{11}(t)&M_{12}(t)\\
        M_{21}(t)&M_{22}(t)\\
    \end{bmatrix}
    \begin{bmatrix}
        C_{n}(0)\\
        D_{n}(0)
    \end{bmatrix},
\nonumber \\ &
n=0,1,2,\dots, 
\end{align}
where
\begin{equation}
\begin{split}\label{C_nD_n}
M_{11}(t) &=\cos\left(\frac{\beta_n t}{2} \right) -\mathrm{i}\frac{\Delta + \chi(2n+1)}{\beta_n}   \sin\left(\frac{\beta_n t}{2} \right),\\ 
M_{12}(t) &=-\mathrm{i}\frac{ 2g \sqrt{n+1}}{\beta_n} \sin\left(\frac{\beta_n t}{2}\right),
\\  M_{22}(t)&= M_{11}^*(t), \quad M_{21}(t)=M_{12}(t), \quad   n=0,1,2,\dots.
\end{split}
\end{equation}
The quantities $|C_n(0)|^2$ and $|D_n(0)|^2$ determine the initial distribution of photons in the excited and ground states of the atom, respectively. Meanwhile, $\beta_n$ is the generalized Rabi frequency caused by AC Stark shifts; these shifts correspond to variations in the energy of an atom resulting from the presence of a non-resonant electric field. The expression for $\beta_n$ is
\begin{equation} \label{Rabi_G}
\beta_n= \sqrt{\left[ \Delta + \chi (2n+1) \right]^{2} + 4g^2 (n+1) }, \quad   n=0,1,2,\dots.
\end{equation}
Once the initial atom-field condition $\ket{\Psi(0)}$ is given, it is possible to obtain the temporal evolution of any observable of the system. In this case, we focus on atomic inversion, $W(t) = \braket{\Psi(t)|\hat{\sigma}_{z}|\Psi(t)}$, which determines the atomic population changes and contains the statistical signature of the field. Thus, the probability of the atom being in its excited state minus the probability of it being in the ground state is determined by the expression
\begin{equation}\label{InversionW}
W(t) = \sum_{n=0}^{\infty} \left(\vert C_n (t) \vert^{2} - \vert D_n (t) \vert^{2} \right).
\end{equation}
Using the solution given in \eqref{0050} and substituting the values of the coefficients given in \eqref{C_nD_n}, we obtain
\begin{align}\label{0100}
W(t)  = & \sum_{n=0}^{\infty}\frac{1}{\beta_n^2}\left\lbrace \left[ \Delta +\left( 2 n+1\right)\chi\right]^2+4g^2 (n+1) \cos \left(\beta_n t \right)\right\rbrace \nonumber 
    \\ 
    & \times \left(\left|C_n(0)\right|^2-\left| D_n(0)\right|^2\right) -\sum_{n=0}^{\infty}\frac{4g \sqrt{n+1}}{\beta_n^2} \nonumber \\ 
    & \times\left[ \Delta +\left( 2 n+1\right)\chi\right] \left[\cos \left(\beta_n t\right)-1\right] C_n(0) D_n(0).
\end{align}
If we suppose that the atom is initially in its excited state, that is, $\ket{\Psi(0)}=\sum_{n=0}^{\infty} C_n (0) \ket{n} \ket{e}$, ($D_n(0)=0$ for $n=0,1,2,\dots$), we can obtain the atomic inversion as follows
\begin{equation}\label{W_e}
\begin{split}
     W_\textrm{e} (t) = \sum_{n=0}^{\infty}\frac{P_n}{\beta_n^2}
\bigg\{&\left[ \Delta+(2n+1)\chi\right]^2\\ &+ 4g^2 (n+1)\cos (\beta_{n} t)  \bigg\},
\end{split}
\end{equation}
where we identify $\left|C_n(0)\right|^2=P_n$ for $n=0,1,2,\dots$, and $P_n$ represents the photon probability distribution. \\
If we now consider that the atom is initially in its ground state, that is, $\ket{\Psi(0)}=\sum_{n=0}^{\infty} D_n (0) \ket{n+1} \ket{g}$ ($C_n(0)=0$ for $n=0,1,2,\dots$), the atomic inversion is given by
\begin{equation}\label{W_g}
 \begin{split}
W_\textrm{g} (t)=  - \sum_{n=0}^{\infty}\frac{P_{n+1}}{\beta_n^2}
\bigg\{&\left[ \Delta+(2n+1)\chi\right]^2\\ &+ 4g^2 (n+1)\cos (\beta_{n} t)  \bigg\},
\end{split}
 \end{equation}
where we now identify $|D_n(0)|^2=P_{n+1}$ for $n=0,1,2,\dots$. This identification makes physical sense: if we analyze the expression~\eqref{Solucion}, which gives us the wavefunction of the complete system, we realize that from the beginning we have assumed that there is one quantum of energy, and therefore, if the atom is in the ground state, the probability of having zero photons in the field is null.

One way to analyze the possible variations of the transition probabilities between the ground state and the first excited state as a function of detuning is by using line shapes, which do not depend on the interaction time duration $t$. We focus on the average atomic inversion $\overline{W}(\Delta)$~\cite{Moya_1991}
\begin{equation} \label{W_Delta}
\overline{W}(\Delta) = \lim_{T \to{\infty}} \frac{1}{T} \int_{0}^{T} W(t) , dt.
\end{equation}
Since
\begin{equation}
\lim_{T \to{\infty}} \frac{1}{T} \int_{0}^{T} \cos (\beta_{n} t) , dt=0,
\end{equation}
and using \eqref{0100}, we have
\begin{align}
\overline{W}(\Delta)  = & \sum_{n=0}^{\infty} \left\lbrace \left[ \frac{\Delta +\left( 2 n+1\right)\chi}{\beta_n}\right]^2
\left(\left|C_n(0)\right|^2-\left| D_n(0)\right|^2\right) \right.
\nonumber \\ 
 & \left. + \; 4g \sqrt{n+1} \left[ \frac{\Delta +\left( 2 n+1\right)\chi}{\beta_n^2}\right]C_n(0) D_n(0) \right\rbrace.
\end{align}
In the case when the atom is initially in the excited state, we use equation~\eqref{W_e} to obtain
\begin{equation}\label{W_Delta_e}
\overline{W}\textrm{e}(\Delta) =\sum_{n=0}^{\infty} P_{n} \left[\frac{\Delta +(2n+1)\chi}{\beta_n}\right]^2,
\end{equation}
while when the atom is initially in the ground state, we use equation~\eqref{W_g} and arrive at the expression
\begin{equation}\label{W_Delta_g}
\overline{W}\textrm{g}(\Delta) =-\sum_{n=0}^{\infty} P_{n+1} \left[\frac{\Delta +(2n+1)\chi}{\beta_n}\right]^2.
\end{equation}

\section{Squeezed coherent states}\label{Comprimidos}
\begin{figure}
\includegraphics{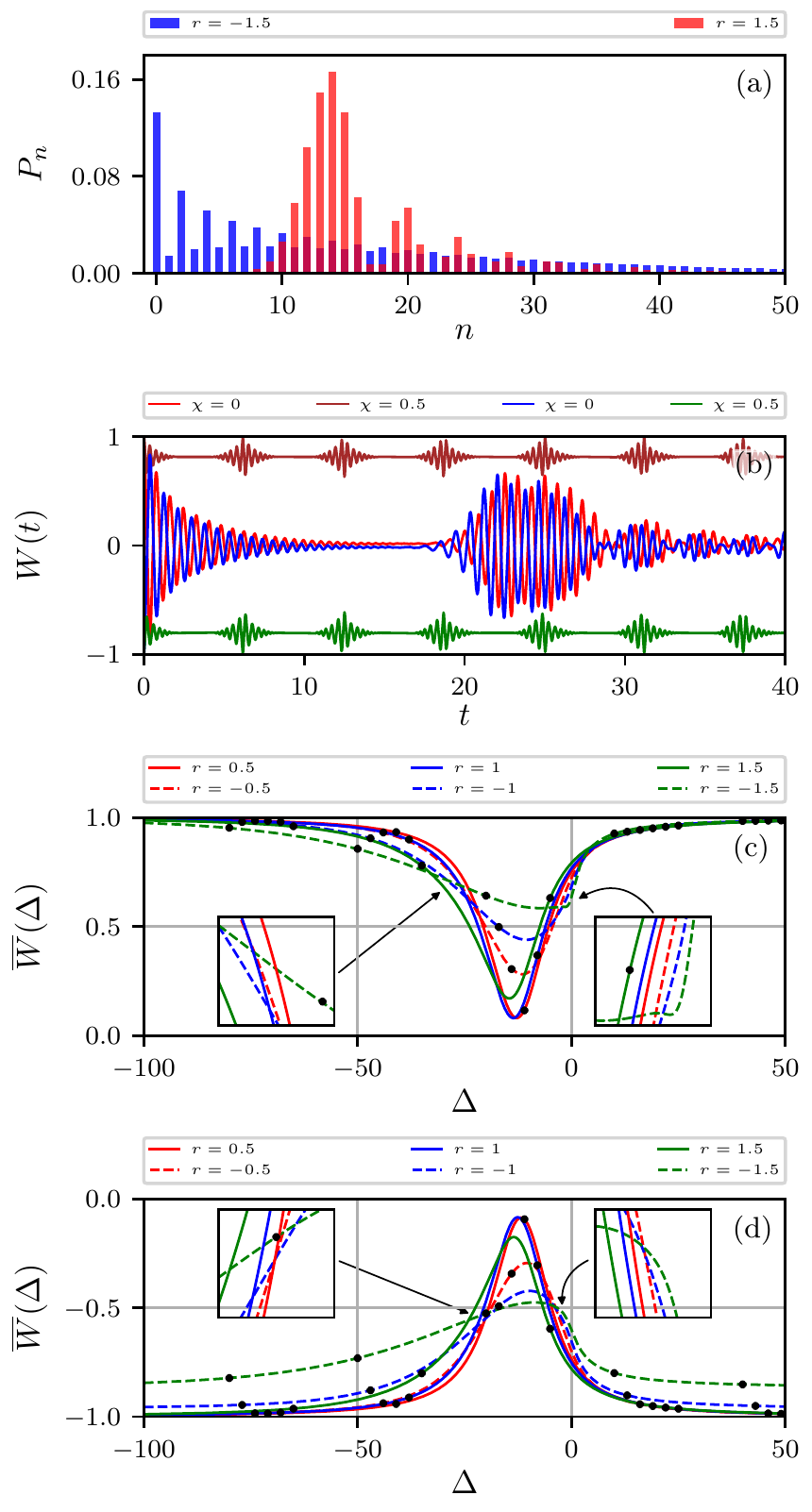}
\centering
\caption{(a)~Photon number distribution for a squeezed coherent state with  $\alpha = 3.5$. (b)~Atomic inversion for $\chi=0$ (red and blue lines) and $\chi=0.5$ (brown and green lines) corresponding to the initial conditions of the atom in the excited and ground state, respectively, with $\Delta = g = 1$. (c)~Average atomic inversion $\overline{W}(\Delta)$ corresponding to the initial condition of the atom in the excited state and the field in a squeezed coherent state, for $\chi=0.5$, $g=1$ and different values of $r$. (d)~In this case, the same situation as before is considered, but preparing the atom in the ground state. The solid and dashed lines correspond to the analytical result, while the dots correspond to the numerical result obtained using Riemann sums in equation \eqref{W_Delta} and QuTiP~\cite{Qutip} to numerically solve the Schrödinger equation.}
\label{fig_2}
\end{figure}

We are interested in studying the case of a compressed coherent field, defined as~\cite[page 155]{Gerry_Book}
\begin{equation} \label{ket1}
    \ket{\alpha,\xi}= \hat{D}(\alpha) \hat{S}(\xi) \ket{0}, 
\end{equation}
where $\hat{D}(\alpha) = \exp (\alpha \hat{a}^{\dagger} - \alpha^{*} \hat{a})$  is the displacement operator, with $\alpha$ a complex number, $\hat{S}(\xi) = \exp [\frac{1}{2} (\xi^{*} \hat{a}^{2} - \xi \hat{a}^{\dagger 2})]$  is the squeeze operator with $\xi = r e^{i\theta}$, and $\ket{0}$ is the vacuum state. If we consider $\xi$ real (i.e., $\theta=0,\; \xi=r$), the squeezed coherent field given by \eqref{ket1} can be expressed as~\cite[page 163]{Gerry_Book}
\begin{equation}\label{ket2}
\begin{split}
\ket{\alpha,r} & = \frac{1}{\sqrt{\cosh (r)}} \exp \left[ -\frac{1}{2} |\alpha|^2 - \frac{\tanh (r)}{2} \alpha^{* 2}\right] \\
& \quad \times \sum_{n=0}^{\infty}  \frac{\tanh^{n/2}(r)}{2^{n/2} \sqrt{n!}}  H_n \left( \alpha \frac{1 + \tanh (r)}{\sqrt{2 \tanh (r)}} \right) \ket{n},
\end{split}
\end{equation}
where $H_n(x)$ are the Hermite polynomials.\\
The probability of finding $n$ photons in the squeezed coherent field is given by \cite[page 163, Eq.(7.81)]{Gerry_Book}
\begin{equation}\label{Pn_Comprimidos}
\begin{split}
P_n & = \frac{|\tanh(r)|^n}{2^n n! \cosh(r)} \exp \left[ -|\alpha|^2 - \frac{\tanh (r)}{2} \left( \alpha^{* 2} + \alpha^{2} \right) \right]\\
    & \quad \times \left|  H_n \left( \alpha \frac{1 + \tanh (r)}{\sqrt{2 \tanh (r)}} \right) \right|^2.
 \end{split}
\end{equation}
In Fig.~\ref{fig_2}~(a), we show the photon number distribution of the initial squeezed coherent field for two different squeezing parameters, and $\alpha = 3.5$. For $r=-1.5$, we obtain a wider distribution of photons, and we can also observe the difference between even and odd contributions. When $r=1.5$, additional smaller contributions can be observed accompanying the main contribution; these additional peaks will be responsible for resonant revivals in atomic inversion~\cite{Satyanarayana_1989}. In Fig.~\ref{fig_2}~(b), we present the atomic inversion $W(t)$ for an atom with $\Delta = g = 1$, initially in the excited or ground state, and a field initially in a squeezed coherent state $\ket{\alpha,r}$ with $\alpha = 3.5$ and $r=1.5$. When $\chi=0$ (represented by red and blue lines for the excited and ground states, respectively), we observe an increase in the collapse time and the appearance of resonant revivals in the atomic inversion. However, considering the non-resonant nearby levels ($\chi=0.5$, represented by brown and green lines), the atomic inversion approaches its initial value on average in the same manner as a coherent state. This is because levels outside of resonance limit the effectiveness of the field in driving transitions out of the initial state~\cite{Moya_1992}. Furthermore, we note that the time for the first resurgence to appear shortens with increasing values of $\chi$, and the oscillations become more pronounced with increasing values of $r$ for both initial states of the atom. It is worth noting that for negative values of $r$, the atomic inversion exhibits irregular behavior with poorly defined rebirths. This behavior resembles the response of a thermal state to the field~\cite{Knight_1982}.
In Fig.\ref{fig_2}~(c), we plot the lineshapes for the initial condition where the atom is in its excited state and the field is in a squeezed coherent state for different values of the squeezing parameter $r$; we use $\chi=0.5$, a value that falls within the validity range of the assumed approximation for the Hamiltonian \eqref{eq:1}~\cite{Villanueva_2020b}, and we have set $g=1$. We can observe that for values of $\Delta>0$, i.e., when the frequency of the atom is higher than that of the field, the lineshapes rapidly decay towards the initial state of the atom, i.e., there are  no transitions between the two levels, regardless of the value of the parameter $r$. On the other hand, for values of $\Delta<0$, we notice that as we decrease the values of $r$, the lineshapes become distorted and approach the initial state of the atom  more rapidly. This is due to the widening of the probability distribution as the values of $r$ decrease, as shown in Fig.~\ref{fig_2}~(a). When we increase positive values of $r$, we observe that for small values of $r$, the lineshapes increase in depth until they reach a minimum point at $r = 0.758$, corresponding to a depth of $0.9304$ (numerically obtained), and then begin to decrease in depth, maintaining an almost symmetrical shape with respect to their lowest point. However, as we further increase the value of $r$, the lineshapes become increasingly distorted and lose their symmetry. This behavior is associated with the field undergoing a transition from superpoissonian statistics ($r<0$) to subpoissonian statistics ($0<r<1.34$), and then back to superpoissonian statistics ($r>1.34$)~\cite{Moya_1992}. This behavior is very different from the case of an initial field in a coherent state, where an increase in the average number of photons causes the lineshapes to widen and shift, while maintaining their symmetry with respect to the lowest point, as shown in~\cite{Moya_1991}. On the other hand, in Fig.~\ref{fig_2}~(d), corresponding to the initial condition of the atom in the ground state and for the same parameter values as in the previous case, it is important to note that the lineshapes are not a mirror image of the case where the atom is in the excited state, due to the presence of an additional excitation between the ground and excited states, as assumed in~\eqref{Solucion}.

\section{Superposition of squeezed coherent states}\label{Superposición}
In the previous section it has been shown that the lineshapes are very sensitive to the sign of the squeezing parameter $r$. In this section, we look at the lineshapes produced by a superposition of squeezed coherent states, but not the usual superposition already studied, where the amplitudes of the constituent states are negative and positive~\cite{Vidiella_1995}, instead we write a superposition of the form
\begin{equation} \label{Superposicion}
    \ket{\psi_{\pm}}=\frac{1}{\mathcal{N}_{\pm}}(\ket{\alpha,r}\pm\ket{\alpha,-r}),
\end{equation}
where $\mathcal{N}_{\pm} = \sqrt{2[1\pm 1/\sqrt{\cosh (2r)}]}$ is the normalization constant.

\begin{figure}
\includegraphics[width = \linewidth]{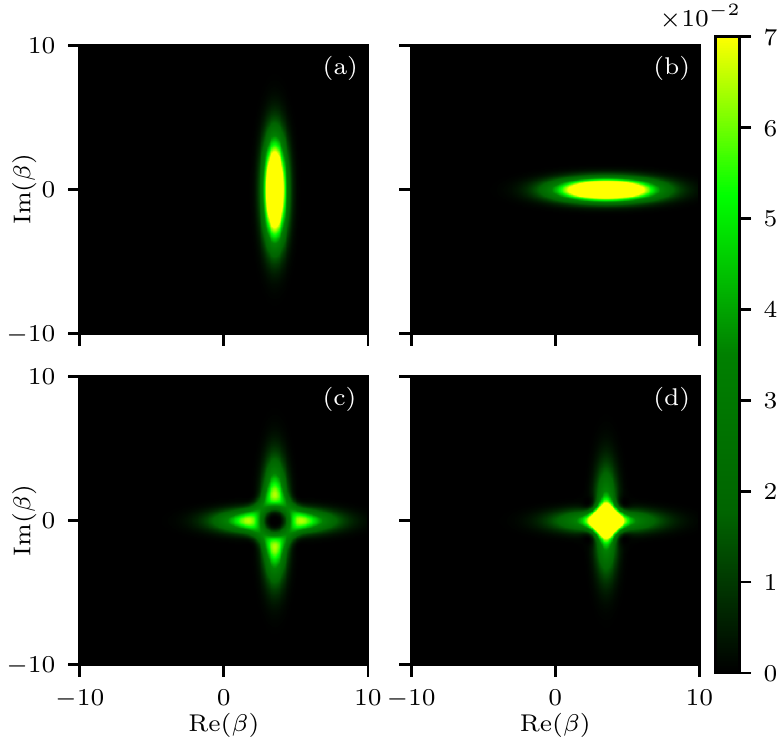}
\centering
\caption{Husimi $Q$-function for the squeezed coherent states with amplitude $\alpha = 3.5$, in (a) $r = 1.5$,  in (b) $r =-1.5$. A superposition of squeezed coherent states of the form (\ref{Superposicion}), in (c) a negative superposition, and in (d) a positive superposition.}
\label{Q_function}
\end{figure}

One way to visualize the behavior of quantum systems in phase space is through the Husimi $Q$-function. This function, also known as the quasiprobability distribution, was introduced by Kôdi Husimi in 1940~\cite{Husimi_1940} and is commonly expressed as the expected value of the density operator~\cite{Moya_Book},
\begin{equation}\label{Husimi}
Q(\beta) = \frac{1}{\pi} \langle \beta | \hat{\rho} | \beta \rangle.
\end{equation}
Based on the results obtained in Appendix~\ref{A_Husimi}, we plot the Husimi $Q$-function for the squeezed coherent states~\eqref{ket1} and for a superposition of the form \eqref{Superposicion} with $\alpha = 3.5$. In Fig.~\ref{Q_function}~(a), we can observe that for positive values of $r$, the squeezing occurs in the real part of $\beta$, i.e., the uncertainty increases in the real part of $\beta$ and decreases in the imaginary part. On the other hand, for a negative squeezing parameter $r=-1.5$, in Fig.~\ref{Q_function}~(b) the  squeezing  occurs in the imaginary part of $\beta$, which implies that the uncertainty increases in the imaginary part and decreases in the real part. In Fig.~\ref{Q_function}(c), corresponding to a negative superposition, we can observe that there is  squeezing in both directions, except in the center. Finally, in Fig.~\ref{Q_function}(d), corresponding to a positive superposition, we can observe from the heat map that there is a maximum  squeezing  only at the center.

Furthermore, the probability of finding $n$ photons in a field defined by the state vector (\ref{Superposicion}) is
\begin{equation} \label{Pn_Superposición}
    P_{n}^{(\pm)}=\frac{1}{\mathcal{N}_{\pm}^2} \left| \braket{n | \alpha,r} \pm \braket{n | \alpha,-r} \right|^2.
\end{equation}
\begin{figure}
\includegraphics{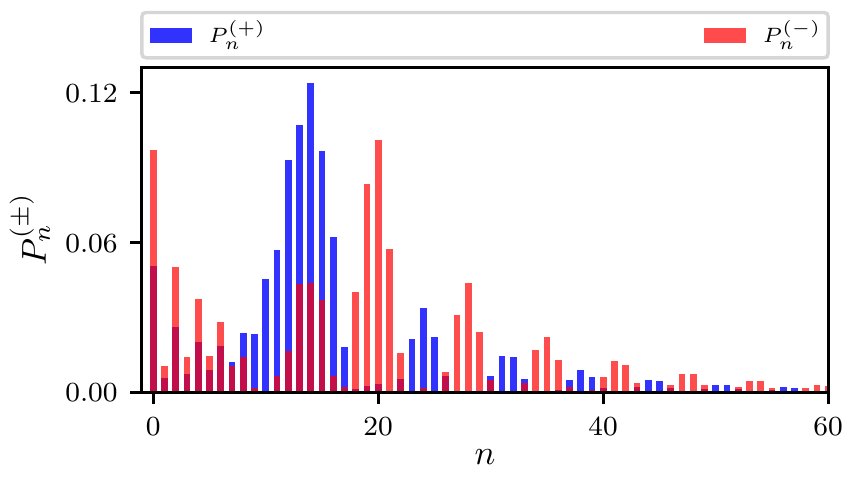}
\centering
\caption{The probability distribution of photons for a superposition of squeezed coherent states $P_{n}^{(\pm)}$, for $\alpha= 3.5$ and $r=1.5$.}
\label{Pn_Superposition}
\end{figure}
In Fig.~\ref{Pn_Superposition}, we show the photon number distribution for a squeezed coherent state superposition, \eqref{Superposicion}, with $r=1.5$ and $\alpha=3.5$. In the superposition $\ket{\psi_+}$, we observe a photon number distribution $P_n^{(+)}$ that resembles that of a squeezed coherent state, with small contributions accompanying the main distribution. However, now $P_n^{(+)}$ also presents contributions before the main contribution. On the other hand, the photon number distribution $P_n^{(-)}$ for the superposition $\ket{\psi_-}$ is notably different from that of $\ket{\psi_+}$, since $P_n^{(-)}$ presents more than one main contribution, as well as a broader distribution.

\begin{figure}[H]
\includegraphics{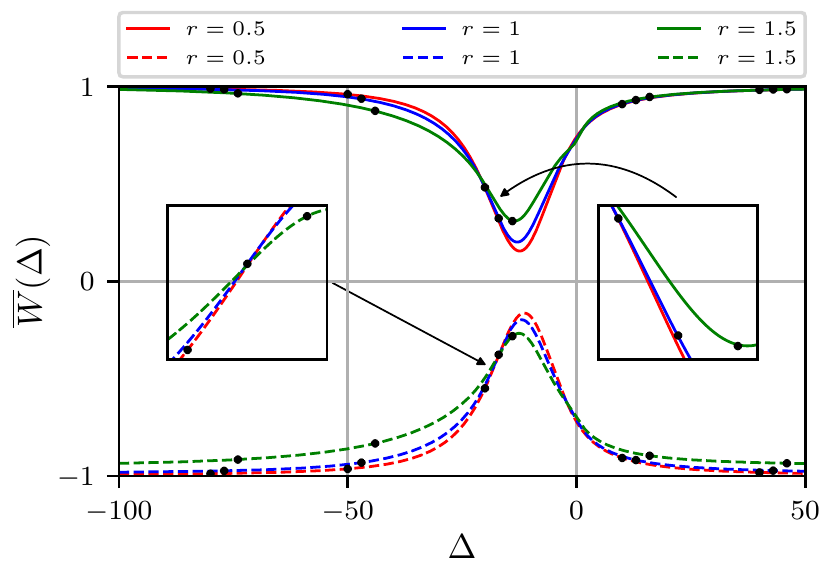}
\centering
\caption{Average atomic inversion $\overline{W}(\Delta)$ for the atom initially in the excited state (solid lines) and in the ground state (dashed lines), and the field in a superposition of squeezed coherent states $P_{n}^{(+)}$, for $\chi=0.5$, $g=1$, $\alpha=3.5$, and different values of $r$. The solid and dashed lines correspond to the analytical result, while the dots correspond to the numerical results.}
\label{Super_Positiva}
\end{figure}

We study now the average atomic inversion, i.e., the lineshapes, when the initial state of the field is a superposition of squeezed coherent states $P_{n}^{(\pm)}$. 
The following image, Fig.~\ref{Super_Positiva},  shows the lineshapes associated with a superposition of squeezed coherent states $\ket{\psi_+}$, with different initial conditions of the atom and for various values of $r$; in that plot employs the following parameter values: $\chi=0.5$, $g=1$, and $\alpha=3.5$. We observe that, when the atom is initially in the excited state (solid lines), for values of $\Delta>0$, the lineshapes decay rapidly towards the initial state of the atom, indicating no transitions between the two levels, regardless of the value of the parameter $r$. In contrast, for values of $\Delta<0$, as we increase the values of $r$, the lineshapes for the superposition $\ket{\psi_+}$ exhibit behavior very similar to that of a squeezed coherent state with $r>0$ (see Fig.~\ref{fig_2}~(c)) due to the similarity of their photon distributions. Specifically, the lineshapes increase in depth until reaching a minimum point at $r=0.308$, corresponding to a depth of $0.8478$ (numerically obtained), and then start to decrease in depth. However, in this case, the lineshapes have a smaller depth than that of a squeezed coherent state, distort more rapidly as the value of $r$ increases, and the symmetry with respect to their lowest point is less noticeable compared to the case of the squeezed coherent state. Similarly, when the atom is initially in the ground state (dashed lines), the lineshapes resemble those of a squeezed coherent state with $r>0$ (see Fig.~\ref{fig_2}~(d)). However, it is important to reiterate that the lineshapes are not a mirror image of the case where the atom is in the excited state. As we can observe, this distinction becomes more prominent for sufficiently large values of $r$.

\begin{figure}[H]
\includegraphics{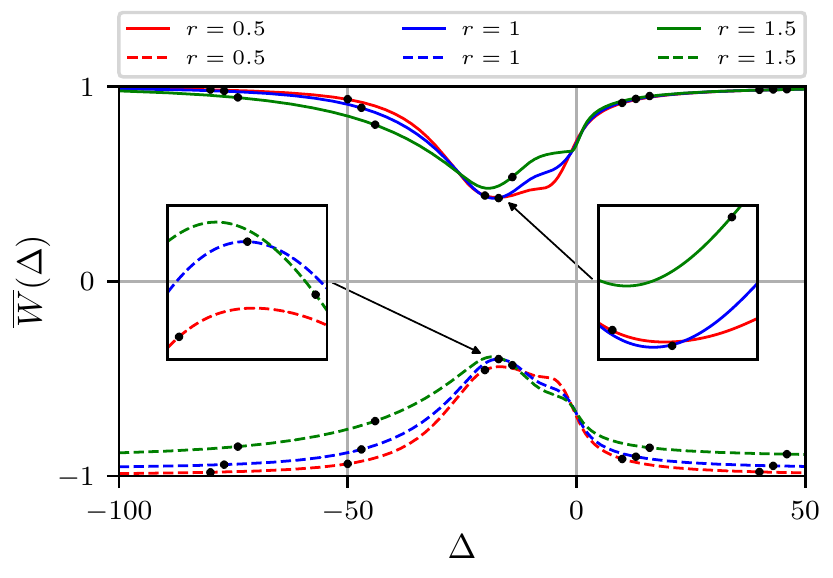}
\centering
\caption{Average atomic inversion $\overline{W}(\Delta)$ for the atom initially in the excited state (solid lines) and in the ground state (dashed lines), and the field in a superposition of squeezed coherent states $P_{n}^{(-)}$, for $\chi=0.5$, $g=1$, $\alpha=3.5$, and different values of $r$. The solid and dashed lines correspond to the analytical result, while the dots correspond to the numerical results.}
\label{Super_Negativa}
\end{figure}

Now we analyze the average atomic inversion corresponding to a superposition of squeezed coherent states $\ket{\psi_-}$, with the same atomic conditions as in the previous case (Fig.~\ref{Super_Positiva}) and for the same parameter values. We observe in Fig.~\ref{Super_Negativa} that when the atom is initially in the excited state (solid lines) for values of $\Delta < 0$, the curves decay rapidly in a similar manner to $\ket{\psi_+}$. However, for values of $\Delta > 0$, they exhibit a very different behavior. In this case, the curves decay more slowly but distort rapidly as the values of $r$ increase, and they also display multiple minima. In fact, the number of minima increases as the parameter $\chi$ value increases. However, as mentioned before, the Hamiltonian\eqref{eq:1} does not allow values greater than 1. It is important to note that in this case, the symmetry with respect to the minimum point no longer exists. Finally, when the atom is initially in the ground state (dashed lines), the curves exhibit a much more noticeable difference than in all previous cases. This is due to the superpoissonian shape of its probability distribution (see Fig.~\ref{Pn_Superposition}) and the presence of an additional excitation between the ground and excited states, as mentioned before.

\section{Conclusions}\label{Conclusión}
In summary, we demonstrated that the lineshapes of the micromaser are highly sensitive to the squeezing parameter $r$. When considering small, positive values of $r$, the photon distribution is subpoissonian and the lineshapes of squeezed coherent states are almost symmetric with respect to their minimum point. However, for negative values of $r$, the photon distribution becomes superpoissonian, and the lineshapes distort rapidly, resulting in a broadened and asymmetric profile. This behavior arises due to the two different kinds of photon distributions that are present in squeezed states with negative $r$: an oscillatory and a coherent-like distribution, which are imprinted on opposite sides of the lineshape.

Moreover, we investigated the behavior of superpositions of squeezed coherent states with equal amplitude. When the sign of the superposition is positive, the lineshapes behave similar to that of a squeezed coherent state with $r>0$, although they distort more rapidly. On the other hand, when the sign of the superposition is negative, the lineshapes distort even more rapidly as the values of $r$ increase, and show more than one minimum point. These findings highlight the importance of carefully considering the sign of the squeezing parameter when analyzing the lineshapes of micromaser systems.

\section*{Acknowledgments}
L. Hernández Sánchez thanks the Instituto Nacional de Astrófisica, Óptica y Electrónica (INAOE) and the Consejo Nacional de Ciencia y Tecnología (CONACyT) by the PhD scholarship awarded (No. CVU: 736710).

\appendix

\section{The Husimi $Q$-function}\label{A_Husimi}
The Husimi $Q$-function for a squeezed coherent state $\ket{\psi}=\ket{\alpha,r}$ is obtained from equation~\eqref{Husimi},
\begin{equation*}
Q (\beta)  = \frac{1}{\pi} \langle \beta | \hat{\rho} | \beta \rangle = \frac{1}{\pi} \left| \braket{\beta | \alpha,r} \right|^{2}.
\end{equation*}
Since a coherent state can be written as $\ket{\beta} = \hat{D}(\beta) \ket{0}$~\cite{Gerry_Book,Moya_Book},  which implies that  $\bra{\beta} = \bra{0} \hat{D}^{\dagger}(\beta)$, and a squeezed coherent state as $\ket{\beta, r} = \hat{D}(\beta) \hat{S}(r) \ket{0}$, we get
\begin{equation*}
    Q (\beta)  = \frac{1}{\pi} |\bra{0}  \hat{D}^{\dagger} (\beta) \hat{D} (\alpha) \hat{S}(r) \ket{0} |^{2}.
\end{equation*}
Given that $\hat{D}^{\dagger} (\beta) \hat{D} (\alpha) = \exp [-\mathrm{i} \text{Im}(\beta \alpha^*)] \hat{D}(\alpha -\beta)$ \cite{Gerry_Book,Moya_Book},
\begin{align}
    Q (\beta) & = \frac{1}{\pi} |\bra{0} \hat{D} (\alpha -\beta) \hat{S}(r) \ket{0} |^{2}  \nonumber \\
           & = \frac{1}{\pi} |\braket{0 |\alpha - \beta,r}|^{2}. \nonumber 
\end{align}
Finally,  using equation~\eqref{ket2}, we obtain
\begin{equation}\label{H_comprimido}
    Q (\beta) = \frac{1}{\pi \cosh (r)} \exp\left[ -|\gamma|^2 - \frac{\tanh (r)}{2} \left( \gamma^2 + \gamma^{* 2} \right) \right],
\end{equation}
where $\gamma = \alpha - \beta$.\\
Using similar algebraic techniques, we can derive the Husimi $Q$-function of the superposition of squeezed coherent states given by \eqref{Superposicion},
\begin{align} \label{eq:B2}
 Q (\beta) & =\frac{1}{\pi \mathcal{N}_{\pm}^2} |\braket{\beta | \alpha,r} \pm \braket{\beta | \alpha,-r} |^{2} \nonumber \\
           & = \frac{2\exp\left(-|\gamma|^2\right)}{\pi \mathcal{N}_{\pm}^2 \cosh (r)}  \left[ \cosh \left( \frac{\tanh (r)}{2} (\gamma^2 + \gamma^{* 2}) \right) \right. \nonumber \\
           & \quad \pm \left. \cosh \left( \frac{\tanh (r)}{2} (\gamma^2 - \gamma^{* 2}) \right) \right],
\end{align}
where $\mathcal{N}_{\pm} = \sqrt{2[1\pm 1/\sqrt{\cosh (2r)}]}$ is the normalization constant and $\gamma = \alpha - \beta$.

%
    
\end{document}